\documentclass[conference,letterpaper]{IEEEtran}

\usepackage{amsmath,amssymb,amsfonts,mathtools}
\usepackage{cite}
\usepackage{graphicx}
\usepackage{bm}
\usepackage{booktabs}
\newcommand{\R}{\mathbb{R}}
\newcommand{\ml}{\mathrm{ML}}
\newcommand{\euc}{\mathrm{Euc}}
\newcommand{\avg}{\mathrm{avg}}
\newcommand{\iso}{\mathrm{iso}}
\newcommand{\argmin}{\mathop{\mathrm{argmin}}}

\newtheorem{proposition}{Proposition}
\newtheorem{lemma}{Lemma}

\title{Fourier-Curve Constellations under Tangential Perturbation: Covariance-Aware Soft Demapping on Coded Links}

\author{
\IEEEauthorblockN{Bin~Han\IEEEauthorrefmark{1},
Muxia~Sun\IEEEauthorrefmark{2},
H.~Vincent~Poor\IEEEauthorrefmark{3},
and Hans~D.~Schotten\IEEEauthorrefmark{1}\IEEEauthorrefmark{4}}
\IEEEauthorblockA{\IEEEauthorrefmark{1}RPTU University Kaiserslautern-Landau, Germany,\quad
\IEEEauthorrefmark{2}Beijing Huairou Laboratory, China\\
\IEEEauthorrefmark{3}Princeton University, USA,\quad
\IEEEauthorrefmark{4}German Research Center for Artificial Intelligence (DFKI), Germany}
}

\begin{document}
\maketitle

\begin{abstract}
A Fourier-curve constellation places $M$ points on a closed curve through $k$ complex slots. A Gaussian perturbation along the curve's tangent, whether injected as artificial noise or arising from first-order jitter of the curve parameter, gives every symbol an observation with a symbol-dependent rank-one covariance, and the maximum-likelihood symbol metric differs from the Euclidean rule by one rank-one correction per candidate. We realize this metric as a max-log soft demapper beside a Euclidean correlator bank at $2kM$ additional multiply--accumulate operations per symbol. On a regular $(3,6)$ LDPC-coded link at $(k,M){=}(20,64)$ it recovers $5.1$\,dB of the Euclidean mismatch at BLER${=}10^{-1}$ under natural labeling and $1.1$\,dB under Gray labeling, of which an average-covariance receiver recovers $0.7$\,dB and nothing measurable, respectively, and it makes the tangential perturbation $0.2$ to $1.0$\,dB cheaper than white noise of the same power on the same codebook. The per-tone phase orientation of the curve acts as an orthogonal rotation, so these results hold at every orientation, and the demapper decodes at the level of an exactly oriented receiver up to $0.2$\,rad of orientation error per component. A bit-interleaved coded-modulation achievable rate corroborates the ordering, a Woodbury extension keeps the rank-one structure under per-tone Rician fading, and $6$-bit lookup-table quantization costs no measurable degradation.
\end{abstract}

\begin{IEEEkeywords}
Fourier-curve constellation, tangential perturbation, covariance-aware demapping, mismatched decoding, bit-interleaved coded modulation, LDPC.
\end{IEEEkeywords}

\section{Introduction}\label{sec:intro}

A Fourier-curve constellation samples $M$ points from a closed curve through $k$ complex baseband slots, so that each symbol is a point in $\R^{2k}$ and the curve's tangent at every point follows in closed form from a length-$k$ vector of per-tone phases~\cite{han2026letter}. This paper studies the coded reception of such a constellation when the transmitted point is displaced along the local tangent by a Gaussian amount. Two mechanisms produce this displacement. The companion letter~\cite{han2026letter} injects it deliberately as artificial noise (AN), in the tradition of AN for physical-layer security~\cite{goel2008}. First-order jitter of the curve parameter produces the same observation: a symbol sent at $\theta_i+\epsilon$ lands at $\bm{x}(\theta_i)+\epsilon\,\partial\bm{x}/\partial\theta$, and the constant speed of the Fourier curve makes this a Gaussian step along the unit tangent. At $k{=}1$ the constellation is $M$-PSK and the perturbation is phase jitter: every received cloud is elongated along the circle, in a direction that rotates with the transmitted symbol, and the ML decision boundaries bend to follow the clouds while the Euclidean bisectors stay straight, as Fig.~\ref{fig:schematic} shows at $k=1$ and $k=2$. Detectors and constellations designed for strong phase noise exploit the same elongation~\cite{colavolpe2005,krishnan2013}. The receiver problem is the same under either mechanism.

The receiver-side effect of a tangential perturbation is a covariance \emph{label}. Under white noise the observation covariance is the same for every candidate, so the Euclidean nearest-neighbor receiver is maximum likelihood (ML) once the candidate means are known. A tangential perturbation attaches to each candidate a rank-one, symbol-dependent covariance, and the ML rule that reads this label differs from the Euclidean rule by a single rank-one correction per candidate. The gap between the two rules is a mismatched-decoding gap~\cite{ganti2000,scarlett2020,martinez2008}, which the letter~\cite{han2026letter} quantifies for uncoded detection. This paper asks how much of the label a coded bit-interleaved coded-modulation (BICM) receiver recovers, at what implementation cost, and what the perturbation costs against the unperturbed link and against white noise of the same power.

The demapper needs the per-tone phase orientation of the curve, the multi-tone analogue of the carrier phase that a PSK receiver recovers before detection. A known orientation is an orthogonal rotation of the signal set, so the coded link performs identically at every orientation, and we measure how much orientation error the demapper tolerates. The acquisition of the orientation from data-cancelling moments of the received tones, and its consequences when the orientation serves as a secret key, are treated in a companion paper~\cite{han2026keyreuse}.

The letter~\cite{han2026letter} contains the uncoded detection theory, restated here as~\eqref{eq:fourier-curve}--\eqref{eq:covariance}, \eqref{eq:ml-metric}, \eqref{eq:euc-metric}, and Proposition~\ref{prop:pep}. The contributions of this paper are (i)~a max-log soft-demapper realization of the matched rule as a rank-one correction beside a Euclidean correlator bank, with operation and storage counts; (ii)~a regular $(3,6)$ LDPC-coded evaluation under natural and Gray labelings that separates mismatch recovery, the residual over an average-covariance receiver, and the cost of the perturbation against the unperturbed link and against white noise of equal power, corroborated by a BICM achievable-rate (AIR) computation; (iii)~the orientation-equivalence lemma and the measured orientation-error tolerance of the demapper; and (iv)~robustness evidence under $6$-bit lookup-table (LUT) quantization, $(\sigma_c,\beta)$ mismatch, and a Woodbury extension to per-tone Rician fading.

The remainder of the paper is organized as follows. Sec.~\ref{sec:model} states the model, the three symbol metrics, and the orientation lemma; Sec.~\ref{sec:results} reports the coded, achievable-rate, robustness, orientation-tolerance, fading, and complexity results; Sec.~\ref{sec:conclusion} concludes.

\section{System Model}\label{sec:model}

\subsection{Constellation, perturbation, and covariance structure}

The transmitter encodes an information block with a binary LDPC code and maps the coded bits to symbols from a codebook of $M$ points in $\R^{2k}$; each symbol occupies $k$ complex baseband slots, written throughout in the real-equivalent domain (slot $j$'s in-phase/quadrature pair forms components $2j{-}1,2j$). The codebook is oriented by a phase vector $\bm{\varphi}=(\varphi_1,\ldots,\varphi_k)\in[0,2\pi)^k$, the per-tone phase offsets the receiver sees after equalization. Following~\cite{han2026letter}, the codebook samples the Fourier curve $\bm{x}(\theta;\bm{\varphi})$ at the $M$ angles $\theta_i=2\pi(i-1)/M$:
\begin{equation}
\begin{split}
\bm{x}_i(\bm{\varphi})=\tfrac{1}{\sqrt{k}}\bigl(&\cos(\theta_i{+}\varphi_1),\sin(\theta_i{+}\varphi_1),\ldots,\\
&\cos(k\theta_i{+}\varphi_k),\sin(k\theta_i{+}\varphi_k)\bigr)
\end{split}
\label{eq:fourier-curve}
\end{equation}
in $\R^{2k}$. The derivative of $\bm{x}(\theta;\bm{\varphi})$ with respect to the curve parameter $\theta$ has constant norm $\|\partial\bm{x}/\partial\theta\|\equiv v_k=\sqrt{(k+1)(2k+1)/6}$, and the unit tangent at the $i$-th codebook point is
\begin{equation}
\begin{split}
\hat{\bm{t}}_i(\bm{\varphi})=\tfrac{1}{\sqrt{k}\,v_k}\bigl(&{-}\sin(\theta_i{+}\varphi_1),\cos(\theta_i{+}\varphi_1),\\
&{-}2\sin(2\theta_i{+}\varphi_2),2\cos(2\theta_i{+}\varphi_2),\ldots,\\
&{-}k\sin(k\theta_i{+}\varphi_k),k\cos(k\theta_i{+}\varphi_k)\bigr).
\end{split}
\label{eq:tangent}
\end{equation}
Constant speed makes the perturbation energy independent of the symbol, and both ends read $\hat{\bm{t}}_i$ from a lookup table.

The orientation is constant over each LDPC block. The codebook is invariant under the joint shift $(\varphi_1,\ldots,\varphi_k)\!\to\!(\varphi_1{-}2\pi/M,\ldots,\varphi_k{-}2k\pi/M)$, which relabels indices $i\mapsto i{-}1$ cyclically, so orientations in one such orbit produce the same signal set up to an $M$-way label shift, the analogue of the phase ambiguity of $M$-PSK. The parameter $k$ sets the ambient dimension $2k$ and $M$ the bits per symbol $B_{\rm mod}=\log_2 M$; we work in the regime $k<M/2$, where the curve is curved at the scale of the codebook spacing and every harmonic step $2\pi m/M$ stays below $\pi$.

Signals are baseband samples after synchronization and equalization ($k$ subcarriers or time-domain samples per symbol), an idealization that Sec.~\ref{sec:fading} relaxes. When message $s_\ell$ is sent, the received vector is
\begin{equation}
\bm{Y}=\bar{\bm{x}}_{s_\ell}(\bm{\varphi})+\sqrt{\beta}\,z_\ell\,\hat{\bm{t}}_{s_\ell}(\bm{\varphi})+\bm{N},\ \ \bar{\bm{x}}_i(\bm{\varphi})=\sqrt{1-\beta}\,\bm{x}_i(\bm{\varphi}),
\label{eq:channel}
\end{equation}
with $z_\ell\sim\mathcal{N}(0,1)$ independent of $\bm{N}\sim\mathcal{N}(\bm{0},\sigma_c^2\bm{I}_{2k})$, where $\beta\in[0,1)$ is the fraction of the unit average transmit energy carried by the perturbation. Under the jitter reading $\sqrt{\beta}z_\ell=\epsilon_\ell v_k$, so $\beta{=}0.3$ at $k{=}20$ corresponds to a parameter jitter of $0.046$\,rad, about half the codebook spacing $2\pi/M$. Since $\bm{x}_i\perp\hat{\bm{t}}_i$, the transmit energy of a symbol is $1{-}\beta{+}\beta z_\ell^2$, with variance $2\beta^2$ and a Gaussian tail that a peak-power-limited front-end would have to clip. Tone $m$ carries $\beta m^2/(kv_k^2)$ of the perturbation energy on average, whereas the mean allocates $(1{-}\beta)/k$ to every tone. Conditioned on~$i$, $\bm{Y}$ is Gaussian with mean $\bar{\bm{x}}_i$ and rank-one-perturbed covariance
\begin{equation}
\bm{\Sigma}_i=\sigma_c^2\bm{I}_{2k}+\beta\,\hat{\bm{t}}_i\hat{\bm{t}}_i^{\top}.
\label{eq:covariance}
\end{equation}

\begin{figure*}[!t]
\centering
\includegraphics[width=.85\textwidth]{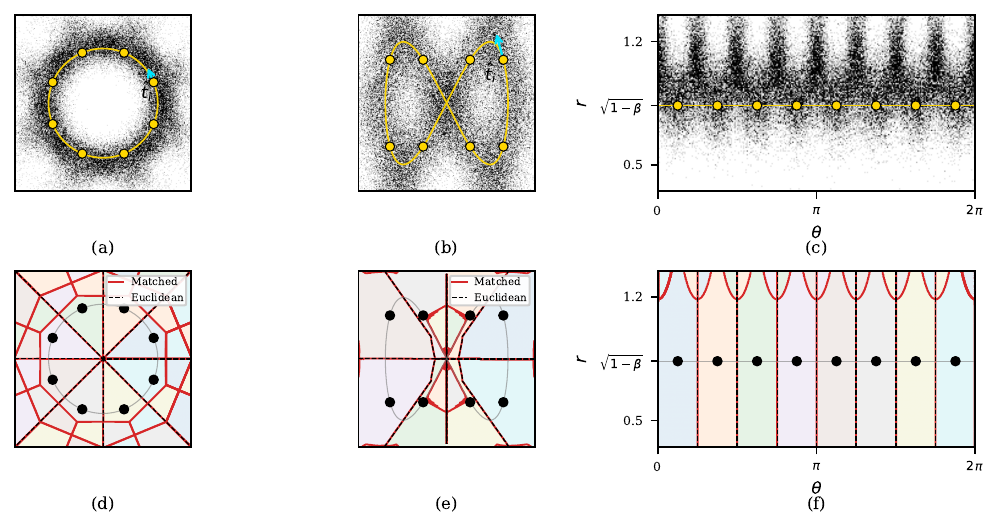}
\caption{Tangential-perturbation geometry at $M{=}8$, $\beta{=}0.3$, $\sigma_c{=}0.14$: received densities with $\hat{\bm{t}}_i$ (top) and matched (red) vs.\ Euclidean (dashed) decision regions (bottom), at $k{=}1$ (a,d), at $k{=}2$ projected to $(\cos\theta,\sin 2\theta)/\sqrt{2}$ (b,e; regions here are projection artefacts), and in intrinsic $(\theta,r)$ coordinates (c,f).}
\label{fig:schematic}
\end{figure*}
In the intrinsic $(\theta,r)$ coordinates of Fig.~\ref{fig:schematic}, the perturbation spreads the clouds along $r{=}\sqrt{1-\beta}$, and the matched regions widen along $\theta$ while the Euclidean bisectors stay vertical.

\subsection{Matched, Euclidean, and average-covariance decoders}\label{sec:decoders}

Substituting~\eqref{eq:covariance} into the ML rule, which minimizes $(\bm{y}-\bar{\bm{x}}_i)^{\top}\bm{\Sigma}_i^{-1}(\bm{y}-\bar{\bm{x}}_i)+\log\det\bm{\Sigma}_i$, the log-det term drops because $\det\bm{\Sigma}_i=\sigma_c^{2(2k-1)}(\sigma_c^2+\beta)$ is invariant in~$i$, and the Sherman--Morrison identity gives
\begin{equation}
\bm{\Sigma}_i^{-1}=\frac{1}{\sigma_c^2}\bm{I}_{2k}-\frac{\beta}{\sigma_c^2(\sigma_c^2+\beta)}\hat{\bm{t}}_i\hat{\bm{t}}_i^{\top}.
\label{eq:sherman-morrison}
\end{equation}
Writing $\bm{r}_i=\bm{y}-\bar{\bm{x}}_i$, the matched ML rule of~\cite{han2026letter} is
\begin{equation}
\widehat{i}_{\ml}
=\argmin_{1\leqslant i\leqslant M}
\biggl\{
\frac{\|\bm{r}_i\|^2}{\sigma_c^2}
-\frac{\beta\,(\bm{r}_i^{\top}\hat{\bm{t}}_i)^2}{\sigma_c^2(\sigma_c^2+\beta)}
\biggr\},
\label{eq:ml-metric}
\end{equation}
while the Euclidean-mismatched decoder retains only the first term:
\begin{equation}
\widehat{i}_{\euc}=\argmin_{1\leqslant i\leqslant M}\|\bm{r}_i\|^2/\sigma_c^2 .
\label{eq:euc-metric}
\end{equation}
We write $\Lambda_{\ell,i}$ for the bracketed metric at symbol slot $\ell$ and candidate $i$ under either rule; the $1/\sigma_c^2$ scale in~\eqref{eq:euc-metric} is immaterial to the hard decision and fixes the Euclidean LLR scale used below. A third rule isolates the two ingredients of the correction. Averaging~\eqref{eq:covariance} over equiprobable symbols, the auxiliary input model of the BICM receiver, gives the candidate-independent matrix $\bm{C}_{\avg}=\sigma_c^2\bm{I}_{2k}+\tfrac{\beta}{M}\sum_i\hat{\bm{t}}_i\hat{\bm{t}}_i^{\top}$, which for the uniform codebook at $k<M/2$ is diagonal with entry $\sigma_c^2+\beta m^2/(2kv_k^2)$ on both real slots of tone $m$ by root-of-unity orthogonality. The \emph{average-covariance} decoder
\begin{equation}
\widehat{i}_{\avg}=\argmin_{1\leqslant i\leqslant M}\bm{r}_i^{\top}\bm{C}_{\avg}^{-1}\bm{r}_i
\label{eq:avg-metric}
\end{equation}
keeps the candidate means and the per-tone coloring of the perturbation and discards the candidate-specific orientation of the covariance. Its difference to~\eqref{eq:euc-metric} measures the static coloring; its difference to~\eqref{eq:ml-metric} is the residual of the exact candidate-dependent receiver over this one auxiliary receiver.

Decomposing the residual $\bm{r}_i=(\bm{r}_i^{\top}\hat{\bm{t}}_i)\hat{\bm{t}}_i+\bm{r}_{i,\perp}$ into tangent and normal components, the matched metric~\eqref{eq:ml-metric} rewrites as
\begin{equation}
\frac{\|\bm{r}_i\|^2}{\sigma_c^2}-\frac{\beta(\bm{r}_i^{\top}\hat{\bm{t}}_i)^2}{\sigma_c^2(\sigma_c^2+\beta)}=\frac{\|\bm{r}_{i,\perp}\|^2}{\sigma_c^2}+\frac{(\bm{r}_i^{\top}\hat{\bm{t}}_i)^2}{\sigma_c^2+\beta}.
\label{eq:decomp}
\end{equation}
The Euclidean metric weights all residual directions by $1/\sigma_c^2$ and so over-penalizes the tangent direction, whose true per-component variance is $\sigma_c^2+\beta$; the matched metric applies the correct tangent variance~\cite{poor1994}. The pairwise consequence has a closed form.

\begin{proposition}[{\cite[Prop.~1]{han2026letter}}]\label{prop:pep}
For codewords $i\neq j$, let $\delta_{ij}=\|\bm{x}_j-\bm{x}_i\|$ and $\cos\alpha_{ij}=\hat{\bm{t}}_i^{\!\top}(\bm{x}_j-\bm{x}_i)/\delta_{ij}$. Under~\eqref{eq:channel}, the Euclidean decoder~\eqref{eq:euc-metric} errs pairwise with probability
\begin{equation}
P_{\euc}(i\!\to\! j)=Q\!\left(\frac{\sqrt{1-\beta}\,\delta_{ij}}{2\sqrt{\sigma_c^2+\beta\cos^2\alpha_{ij}}}\right).
\label{eq:pep}
\end{equation}
\end{proposition}
The proof projects the perturbation onto the normalized chord, which adds $\beta\cos^2\alpha_{ij}$ to the variance of the Gaussian score difference~\cite{han2026letter}; the penalty enters only through the tangent--chord alignment that~\eqref{eq:decomp} re-weights. As $\sigma_c\!\to\!0$ the argument of~\eqref{eq:pep} tends to $\sqrt{1-\beta}\,\delta_{ij}/(2\sqrt{\beta}\,|\cos\alpha_{ij}|)$, which is finite whenever $\cos\alpha_{ij}\neq0$. For the uniformly sampled codebook the adjacent chord satisfies $\hat{\bm{t}}_i^{\!\top}(\bm{x}_{i+1}-\bm{x}_i)=(kv_k)^{-1}\sum_{m=1}^{k}m\sin(2\pi m/M)$, and at $k<M/2$ every summand is positive, so the Euclidean receiver retains an SNR-independent symbol-error floor bounded below by $P_{\euc}(i\!\to\!i{+}1)$ at $\sigma_c\!\to\!0$. At $(k,M){=}(20,64)$, $\beta{=}0.3$, the limit is $Q(1.13)\approx0.13$ per neighbor, and an exhaustive numerical check over the $63$ competitors finds the limiting argument minimized at the two neighbors.

The coded receiver forms max-log bit metrics from the symbol metric, a matched symbol metric with max-log BICM demapping that differs from an exact bitwise ML demapper; synchronization and channel estimation are assumed ideal.

\subsection{Orientation equivalence}\label{sec:phase}

The orientation enters the signal set only as a rotation. Let $\bm{R}(\alpha)$ be the $2{\times}2$ rotation by $\alpha$ and $\bm{Q}_{\bm{\varphi}}=\mathrm{diag}(\bm{R}(\varphi_1),\ldots,\bm{R}(\varphi_k))$.

\begin{lemma}\label{lem:invariance}
For every $\bm{\varphi}$, $\bm{x}_i(\bm{\varphi})=\bm{Q}_{\bm{\varphi}}\bm{x}_i(\bm{0})$ and $\hat{\bm{t}}_i(\bm{\varphi})=\bm{Q}_{\bm{\varphi}}\hat{\bm{t}}_i(\bm{0})$. Under~\eqref{eq:channel}, and under the post-zero-forcing fading model of Sec.~\ref{sec:fading}, the conditional laws satisfy $P_{\bm{\varphi}}(A\mid i)=P_{\bm{0}}(\bm{Q}_{\bm{\varphi}}^{\top}A\mid i)$, so every decision rule $\delta_{\bm 0}$ at zero orientation has a conjugate $\delta_{\bm{\varphi}}(\bm y)=\delta_{\bm 0}(\bm{Q}_{\bm{\varphi}}^{\top}\bm y)$ with the same error probability, and conjugate bit metrics have the same generalized mutual information; the matched, Euclidean, and average-covariance rules are conjugate families.
\end{lemma}
The identities follow from~\eqref{eq:fourier-curve} and~\eqref{eq:tangent} tone by tone. Since $\bm{Q}_{\bm{\varphi}}$ is orthogonal, $\bm{Q}_{\bm{\varphi}}^{\top}\bm{Y}=\bar{\bm{x}}_{s_\ell}(\bm{0})+\sqrt{\beta}z_\ell\hat{\bm{t}}_{s_\ell}(\bm{0})+\bm{Q}_{\bm{\varphi}}^{\top}\bm{N}$ with $\bm{Q}_{\bm{\varphi}}^{\top}\bm{N}$ distributed as $\bm{N}$; the per-tone matrix $\bm{D}_h$ of Sec.~\ref{sec:fading} is a multiple of the identity on each tone and commutes with $\bm{Q}_{\bm{\varphi}}$. The coded results below therefore hold at every $\bm{\varphi}$.

\section{Simulation Results}\label{sec:results}

\subsection{Setup}\label{sec:setup}

Throughout we take $k{=}20$ slots per symbol, $M{=}64$ points, and perturbation fraction $\beta{=}0.3$. Dimension $k$ trades occupied slots for per-slot energy at net efficiency $\eta=RB_{\rm mod}/k$ bits per complex slot: sweeping $k\!\in\!\{2,8,20\}$ at $M{=}64$ moves the matched BLER${=}10^{-1}$ waterfall from $27.5$ to $7.6$ to $-3.7$\,dB at $\eta{=}1.51$, $0.38$, and $0.15$ bits per slot; $k{=}20$ is where the waterfall drops below $0$\,dB per slot. $M{=}64$ gives $B_{\rm mod}{=}6$ bits per symbol, matching the rate-$1/2$ code to that efficiency; the natural-labeling mismatch gap grows with codebook density, from $0.6$\,dB at $M{=}16$ to at least $14$\,dB at $M{=}256$. The fraction $\beta{=}0.3$ lies just past the knee at which the Euclidean decoder collapses ($\beta{\approx}0.25$) while the matched decoder degrades gradually~\cite{han2026letter}.

The code is a regular $(3,6)$ Gallager LDPC code at $n_c{=}1008$, built from a fixed permutation seed, with a rank-$502$ parity-check matrix, hence $n_{\mathrm{info}}{=}506$ and $R{=}506/1008\!\approx\!0.502$; decoding runs $50$ scaled-min-sum iterations, $\alpha{=}0.8$. No bit interleaver is used: consecutive coded bits form the $B_{\rm mod}$-bit label of each symbol, natural binary unless stated or the binary-reflected Gray code of the index, the better of the two tested labelings; a codeword spans $n_s{=}168$ symbols. The average per-complex-slot SNR
\begin{equation}
\rho_{\rm slot}=1/(2k\,\sigma_c^2)
\label{eq:snr}
\end{equation}
is the x-axis of Fig.~\ref{fig:bler-combined}; it is a total-energy-per-slot normalization. Total-symbol $E_s/N_0{=}k\rho_{\rm slot}$, and $\eta\!\approx\!0.151$\,bits/complex-slot. LLRs are max-log BICM metrics~\cite{caire1998,szczecinski2015}, $\lambda_{\ell,j}=\tfrac{1}{2}\bigl[\min_{i:b_j(i)=1}\Lambda_{\ell,i}-\min_{i:b_j(i)=0}\Lambda_{\ell,i}\bigr]$, on the chosen $\Lambda_{\ell,i}$ (matched, Euclidean, or average-covariance).

Every BLER point runs until $200$ block errors or $10^4$ codewords, whichever comes first, with independent codewords, noise, and perturbation draws per curve. The plotted bands are Clopper--Pearson intervals at the realized sample size; because the sample size is a stopping time, their coverage is approximate and they serve as descriptive uncertainty. Points with zero errors in $10^4$ codewords are reported as BLER${<}3.0{\cdot}10^{-4}$, the one-sided $95\%$ bound $1-0.05^{1/n}$. Waterfalls are the BLER${=}10^{-1}$ crossings, interpolated linearly in $(\rho_{\rm slot}$ in dB$,\log_{10}\mathrm{BLER})$ between the two bracketing grid points; every quoted waterfall difference carries a $95\%$ percentile interval from a parametric bootstrap of the bracket counts, conditional on the realized sample sizes and on the interpolation rule.

The evaluation compares five simulated configurations and one derived one. The \emph{proposed} scheme pairs the perturbed Fourier-curve transmitter with the matched receiver. \textbf{B1} keeps the same transmitter and decodes with the Euclidean rule, and \textbf{B1$^{\avg}$} with the average-covariance rule~\eqref{eq:avg-metric}. \textbf{B2} replaces the constellation by a flat random-spherical codebook on $S^{2k-1}$ (i.i.d.\ Gaussian points normalized to the sphere, index labeling unoptimized for BICM), scaled by $\sqrt{1{-}\beta}$, with a white perturbation of the same energy and Euclidean decoding. \textbf{B3} is the unperturbed Fourier-curve link ($\beta{=}0$) with Euclidean decoding. \textbf{B3$^{\iso}$} is the Fourier-curve link with a white perturbation of energy $\beta$ instead of the tangential one; white Gaussian perturbation plus AWGN is AWGN, so B3$^{\iso}$ at noise level $\sigma_c$ equals B3 at $\sigma_3^2=(\sigma_c^2+\beta/(2k))/(1-\beta)$ exactly, and its crossings are read from B3 through this map. All simulated curves run in one sweep over $\sigma_c\in[0.10,0.40]$ ($15$ points) under one seed base.

\subsection{Coded BLER}\label{sec:bler}

\begin{figure}[!t]
\centering
\vspace*{3pt}
\includegraphics[width=\linewidth]{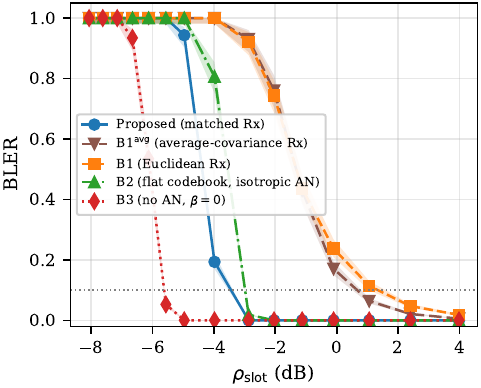}
\caption{Coded BLER vs.\ $\rho_{\rm slot}$ at $(k,M){=}(20,64)$, $\beta{=}0.3$, regular $(3,6)$ LDPC code, natural labeling; bands and configurations as defined in Sec.~\ref{sec:setup}; dotted line at BLER${=}10^{-1}$.}
\label{fig:bler-combined}
\end{figure}

\begin{table}[!t]
\centering
\caption{BLER${=}10^{-1}$ waterfall, its difference $\Delta$ to the proposed receiver (positive: proposed earlier), and AIR crossing of $\eta$, in $\rho_{\rm slot}$ (dB), one run; n.r.: not reached in the tested range. Bootstrap $95\%$ half-widths of $\Delta$: $0.18$\,dB (B1) and $0.12$\,dB (B1$^{\avg}$) under natural labeling, at most $0.04$\,dB otherwise.}
\label{tab:receivers}
\footnotesize
\begin{tabular}{@{}l@{\hskip 5pt}r@{\hskip 5pt}r@{\hskip 5pt}r@{\hskip 8pt}r@{\hskip 5pt}r@{\hskip 5pt}r@{}}
\toprule
 & \multicolumn{3}{c}{Natural} & \multicolumn{3}{c}{Gray} \\
\cmidrule(lr){2-4}\cmidrule(lr){5-7}
Receiver & BLER & $\Delta$ & AIR & BLER & $\Delta$ & AIR \\
\midrule
Proposed & $-3.87$ & -- & $-5.15$ & $-4.88$ & -- & $-5.97$ \\
B1$^{\avg}$ & $+0.53$ & $4.40$ & n.r. & $-3.86$ & $1.03$ & $-4.75$ \\
B1 & $+1.26$ & $5.13$ & n.r. & $-3.83$ & $1.06$ & $-4.38$ \\
B2 & $-3.36$ & $0.51$ & $-4.42$ & $-3.38$ & $1.50$ & $-4.44$ \\
B3$^{\iso}$ & $-3.64$ & $0.23$ & $-4.85$ & $-3.84$ & $1.04$ & $-5.04$ \\
B3 & $-5.72$ & $-1.85$ & $-6.81$ & $-5.90$ & $-1.02$ & $-6.98$ \\
\bottomrule
\end{tabular}
\end{table}

Table~\ref{tab:receivers} collects the waterfalls and AIR crossings under both labelings; Fig.~\ref{fig:bler-combined} shows the natural-labeling curves. Four differences carry distinct meanings. The first is mismatch recovery on the perturbed channel: the matched decoder reaches BLER${=}10^{-1}$ at $-3.87$\,dB, $5.13$\,dB ($95\%$ interval $[4.93,5.28]$) earlier than B1 on the same transmitter under natural labeling; the gap follows from the rank-one correction in~\eqref{eq:ml-metric}. Under Gray labeling the matched decoder crosses at $-4.88$\,dB and B1 at $-3.83$\,dB, a gap of $1.06$\,dB ($[1.02,1.09]$). Gray labeling moves the matched waterfall by $1.01$\,dB and B1's by $5.08$\,dB, because B1's adjacent-symbol confusions then cost single bit errors that the code absorbs. The second difference is the residual over the average-covariance receiver: under natural labeling B1$^{\avg}$ crosses at $+0.53$\,dB, so whitening the per-tone coloring recovers $0.73$\,dB of the $5.13$\,dB, and the exact candidate-dependent receiver retains a residual of $4.40$\,dB ($[4.29,4.52]$) over this auxiliary receiver; under Gray labeling B1$^{\avg}$ crosses within $0.03$\,dB of B1, the coloring recovers nothing measurable, and the residual is $1.03$\,dB ($[0.99,1.06]$). The third difference is the cost of the perturbation itself: the unperturbed link B3 crosses at $-5.72$\,dB under natural and $-5.90$\,dB under Gray labeling, so the perturbation costs the matched link $1.85$\,dB ($[1.81,1.88]$) and $1.02$\,dB ($[1.00,1.04]$). The fourth is the shape of the perturbation at fixed energy: white noise of the same power on the same codebook, B3$^{\iso}$, crosses at $-3.64$ and $-3.84$\,dB, so a tangential perturbation read by the matched receiver is $0.23$\,dB (natural) and $1.04$\,dB (Gray) cheaper than white noise, whereas read by the Euclidean receiver it is $4.9$ and $0.0$\,dB more expensive: the structure of the impairment is an asset or a liability depending on whether the receiver exploits it.

B2's white perturbation makes its Euclidean decoder already ML; the matched link runs $0.51$\,dB ahead of it under natural labeling (ten independent draws spread by $0.13$\,dB), a comparison that bundles the codebook with the perturbation shape; B2's Gray column is an arbitrary relabeling of a random codebook.

\subsection{Achievable-rate corroboration}\label{sec:air}

To check that the coded gaps track a metric-level difference, we compute the BICM achievable rate~\cite{caire1998,martinez2008} from the demapper LLRs,
\begin{equation}
I_{\rm BICM}(s)=B_{\rm mod}-\sum_{j=1}^{B_{\rm mod}}\mathbb{E}\!\left[\log_2\!\bigl(1+e^{-\tilde{b}_{\ell,j}\,s\,\lambda_{\ell,j}}\bigr)\right],
\label{eq:air}
\end{equation}
with $\tilde{b}{=}{+}1$ if the true bit is $0$ and $\tilde{b}{=}{-}1$ otherwise. The scalar $s$ corrects for LLR mis-scaling; a golden-section search over $s\in[10^{-3},50]$, one $s$ for all bit levels, yields $I_{\rm BICM}(s_\star)$~\cite{scarlett2020}, a generalized mutual information and hence an achievable rate for the given bit metric. At $2{\times}10^{5}$ symbols per point, the matched demapper's $I_{\rm BICM}(s_\star)/k$ stays above $\eta$ down to $\rho_{\rm slot}\approx-5.2$\,dB and B2 crosses near $-4.4$\,dB, whereas B1's calibrated max-log rate peaks at $0.076$\,bits/complex-slot and stays below $\eta$ throughout the tested range, as does B1$^{\avg}$'s. B1's small rate reflects the symbol-confusion floor established after Proposition~\ref{prop:pep}, observed at ${\approx}26\%$ here; its max-log LLRs grow overconfident as $\sigma_c\!\to\!0$ and the optimizer runs to the boundary $s{=}10^{-3}$ at the four highest-SNR points. Under Gray labeling all three demappers reach $\eta$, matched at $-6.0$\,dB, B1$^{\avg}$ at $-4.8$\,dB, and B1 at $-4.4$\,dB, consistent with the coded Gray gaps. B3 crosses $\eta$ at $-6.8$ and $-7.0$\,dB and B3$^{\iso}$ at $-4.85$ and $-5.04$\,dB, so at the rate level the perturbation costs $1.7$ and $1.0$\,dB against the unperturbed link and the tangential shape saves $0.3$ and $0.9$\,dB against white noise, in line with the coded numbers.

\subsection{Quantization and parameter-mismatch robustness}\label{sec:quant}

The quantizer is uniform symmetric fixed-point, clipped at $\pm 1$ with round-to-nearest, and applies only to the LUT entries $\bar{\bm x}_i$ and $\hat{\bm t}_i$ shared by both ends; branch-metric and LDPC-LLR arithmetic stay double-precision. With quantized tangent $\bm t_{q,i}$ the exact covariance is $\sigma_c^2\bm I+\beta\bm t_{q,i}\bm t_{q,i}^{\!\top}$, whose inverse carries the denominator $\sigma_c^2(\sigma_c^2+\beta\|\bm t_{q,i}\|^2)$ and whose log-det term $\log(1+\beta\|\bm t_{q,i}\|^2/\sigma_c^2)$ becomes candidate-dependent, and $\bar{\bm x}_{q,i}^{\!\top}\bm t_{q,i}$ is no longer zero, so the projection must use the residual. The implemented demapper projects the residual, keeps the unit-tangent denominator $\sigma_c^2{+}\beta$, and drops the log-det, so the sweep tests the approximate shared-LUT receiver and not the exact ML rule under the quantized covariance. The distortion is small: $\max_i|\|\bm t_{q,i}\|^2-1|$ is $5.5{\cdot}10^{-2}$ at $6$ bits and $3.4{\cdot}10^{-4}$ at $12$, and $\max_i|\bar{\bm x}_{q,i}^{\!\top}\bm t_{q,i}|$ is $1.7{\cdot}10^{-2}$ at $6$ bits. Across $6$--$32$ bits, neither demapper shows measurable BLER degradation at $\sigma_c\in\{0.20,0.24,0.28\}$, spanning the pre-, mid-, and post-waterfall regions; LLR and arithmetic quantization are out of scope.

Parameter mismatch is tested at $(\sigma_c,\beta){=}(0.22,0.3)$, about $1$\,dB above the waterfall: with $\hat{\beta}/\beta$ and $\hat{\sigma}_c/\sigma_c$ each swept over $\{0.5,0.75,1,1.25,1.5\}$ while the other is held at unity, all nine cells show zero errors in $10^{3}$ codewords (one-sided $95\%$ bound ${\leqslant}0.3\%$), which rules out gross sensitivity to the scalar coefficients.

\subsection{Orientation-error tolerance}\label{sec:acquisition}

\begin{figure}[!t]
\centering
\vspace*{3pt}
\includegraphics[width=\linewidth]{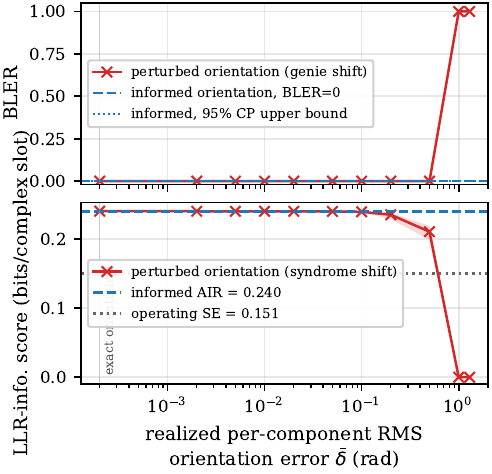}
\caption{BLER (top) and LLR-information score (bottom) of the matched receiver operating at a perturbed orientation, vs.\ the realized per-component RMS orientation error $\bar{\delta}$, at $(k,M){=}(20,64)$, $\beta{=}0.3$, $\sigma_c{=}0.18$. Dashed: the orientation-informed receiver at the same operating point.}
\label{fig:eve-key}
\end{figure}

We measure how much orientation error the demapper tolerates. The receiver is given a perturbed orientation $\tilde{\bm{\varphi}}=(\bm{\varphi}+\bar{\delta}\sqrt{k}\,\bm{u})\bmod 2\pi$ with $\bm{u}$ uniform on the unit sphere of $\R^k$, so that $\bar{\delta}$ is its per-component RMS error, realized as the wrapped-torus distance minimized over the $M$ cyclic equivalences. Its BLER assumes a genie shift selector that counts success if any of the $M$ cyclic relabelings decodes; the lower panel of Fig.~\ref{fig:eve-key} applies~\eqref{eq:air} to the LLRs at the syndrome-weight shift, a block-dependent selection that makes the quantity an empirical LLR-information score and differs from a memoryless BICM rate. Sweeping $\bar{\delta}$ at $\sigma_c{=}0.18$ with $8$ perturbation directions and $2{\cdot}10^{3}$ codewords per direction, pooled to $1.6{\cdot}10^{4}$ codewords per point, the perturbed receiver decodes at the level of the informed receiver up to $\bar{\delta}{=}0.2$; at $\bar{\delta}{=}0.5$ its BLER is $10^{-4}$ and its score ($0.211$ bits/complex-slot) still exceeds $\eta$; at $\bar{\delta}{=}1.0$ decoding collapses (BLER${=}1$, score${\approx}0$). The failure transition lies between $0.5$ and $1.0$\,rad per component, which sets the accuracy that any orientation acquisition, such as the moment-based estimator of~\cite{han2026keyreuse}, has to reach.

\subsection{Per-tone fading and the Woodbury extension}\label{sec:fading}

With per-tone Rician gains $h_j\!\in\!\mathbb{C}$, $E|h_j|^2{=}1$, drawn independently per tone and per codeword and held constant over the $168$ symbols of a codeword, perfect-CSI zero forcing $y_j\!\mapsto\!y_j/h_j$ restores the geometry and inflates the per-tone noise variance to $\sigma_c^2/|h_j|^2$, which a static-$\sigma_c$ demapper mismatches. The post-ZF covariance is still diagonal-plus-rank-one: with $\bm{D}_h$ carrying $|h_j|^{-2}\sigma_c^2$ in both slots of tone $j$, $\bm{\Sigma}_i(\bm{h}){=}\bm{D}_h{+}\beta\hat{\bm{t}}_i\hat{\bm{t}}_i^{\!\top}$, the Woodbury identity and the matrix-determinant lemma give, with $\bm{u}_i{=}\bm{D}_h^{-1}\hat{\bm{t}}_i$ and $\eta_i{=}\hat{\bm{t}}_i^{\!\top}\bm{u}_i=\sum_{m=1}^{k}m^2|h_m|^2/(kv_k^2\sigma_c^2)$,
\begin{equation}
\Lambda_i^{\rm fade}(\bm{y})=\bm{r}_i^{\!\top}\bm{D}_h^{-1}\bm{r}_i-\frac{\beta(\bm{r}_i^{\!\top}\bm{u}_i)^2}{1+\beta\eta_i}+\log(1+\beta\eta_i),
\label{eq:fading-metric}
\end{equation}
where $\eta_i$ is independent of the candidate and of $\bm{\varphi}$, so the log term cancels in candidate comparisons; the metric keeps the $O(kM)$ cost and needs only the per-tone gains from CSI. Sweeping the Rician factor $K\!\in\!\{\infty,100,10,1,0.1,0.01,0\}$ at $\sigma_c{=}0.15$ with one fading draw per codeword, the static-$\sigma_c$ matched decoder maintains BLER${\leqslant}6{\times}10^{-4}$ for $K\!\geqslant\!10$ and saturates near $0.69$ at pure Rayleigh (Euclidean: $0.97$), whereas the Woodbury demapper maintains BLER${\leqslant}2{\times}10^{-4}$ at every tested $K$. The rank-one structure therefore generalizes beyond AWGN under perfect CSI and exact zero forcing.

\subsection{Complexity}\label{sec:complexity}

We count one MAC as one multiplication with accumulation and exclude LDPC decoding and LLR min-search. Per received symbol, the direct Euclidean stage $\|\bm y-\bar{\bm x}_i\|^2$ costs $2kM$ subtractions, $2kM$ MACs, and $2kM$ LUT reads; the correlator form $\bm y^{\!\top}\bar{\bm x}_i$, admissible because $\|\bar{\bm x}_i\|^2{=}1{-}\beta$, costs $2kM$ MACs and $2kM$ LUT reads. The tangent correction adds the same increment to either form: one $2k$-dimensional projection per candidate, $2kM{=}2560$ MACs and $2kM$ LUT reads at $(k,M){=}(20,64)$, followed by $M$ squarings and $M$ scaled subtractions; since $\bar{\bm x}_i^{\!\top}\hat{\bm t}_i{=}0$ the projection can be taken from $\bm y$ directly. The LUT stores $2M\cdot 2k{=}5120$ values: $20$\,KB total at $32$-bit precision, of which the tangent block is the $10$\,KB increment.

\section{Conclusion}\label{sec:conclusion}

A tangential Gaussian perturbation of a Fourier-curve constellation, injected or arising from parameter jitter, gives every symbol a rank-one covariance label, and the receiver that reads it is a rank-one correction beside a Euclidean correlator bank, costing $2kM$ MACs per symbol and $10$\,KB of additional LUT storage. At $(k,M){=}(20,64)$ with regular $(3,6)$ LDPC coding it recovers $5.1$\,dB of the Euclidean mismatch at BLER${=}10^{-1}$ under natural labeling and $1.1$\,dB under Gray labeling, of which whitening the average covariance accounts for $0.7$\,dB and nothing measurable, respectively; the perturbation costs $1.8$ and $1.0$\,dB against the unperturbed link and, read by this receiver, is $0.2$ and $1.0$\,dB cheaper than white noise of the same power. The BICM achievable rate, $6$-bit LUT quantization, scalar mismatch, and a Woodbury extension to per-tone Rician fading confirm the ordering. The demapper tolerates orientation errors up to $0.2$\,rad per component at the informed level and fails between $0.5$ and $1.0$\,rad. Standard codes, peak-power constraints, channel estimation, and front-end impairments are future work; Chase retransmission on this link is analyzed in~\cite{han2026chase}.

\end{document}